\documentclass[prl,aps,onecolumn,superscriptaddress,preprint]{revtex4}
\usepackage[dvips,final]{graphicx}
\usepackage{txfonts}

\newcommand{\const}{\mathrm{const}}
\newcommand{\tr}{\mathop{\mathrm{tr}}}
\newcommand{\trK}{\tr K}
\newcommand{\Mdot}{{\dot M}}

\begin{document}
\title{Transonic and subsonic flows in general relativistic radiation
hydrodynamics}
\author{Janusz Karkowski}
\affiliation{M. Smoluchowski Institute of Physics, Jagiellonian
University,
Reymonta 4, 30-059 Krak\'{o}w, Poland}
\author{Edward Malec}
\affiliation{M. Smoluchowski Institute of Physics, Jagiellonian
University,
Reymonta 4, 30-059 Krak\'{o}w, Poland}
\affiliation{Physics Department, UCC, Cork, Ireland}
\author{Krzysztof Roszkowski}
\affiliation{M. Smoluchowski Institute of Physics, Jagiellonian
University,
Reymonta 4, 30-059 Krak\'{o}w, Poland}
\author{Zdobys\l aw \'Swierczy\'nski}
\affiliation{Pedagogical University, Podchor\c a\.zych 1, Krak\'{o}w,
Poland}

\begin{abstract}
We analyze stationary accretion of selfgravitating gas onto a compact
center within general-relativistic ra\-diation hydrodynamics. Spherical symmetry
and thin gas approximation are assumed. Numerical investigation shows that
transonic flows exist for small redshifts and they cease to exist for
high redshifts and high luminosities. There exist two branches of flows
(subsonic or supersonic) that originate at a bifurcation point and that embrace the set
of subsonic solutions. The morphology of the set of subsonic solutions is
essentially independent of redshifts and flows that belong to their boundary provide
estimates
of the gas abundance of subsonic solutions. It appears that prescribed
boundary
data guarantee uniqueness only of the bifurcation point, and that the
latter
has maximal  luminosity.
\end{abstract}

\maketitle

\section{Introduction.}
Consider a general-relativistic system --- a compact core immersed in a
steadily accreting selfgravitating gas. The gravitational binding energy
of the
infalling gas can be converted to a radiation. Assume that an external
distant
observer can measure total luminosity, asymptotic temperature and
redshifts of
the radiation. Let be known: the total (asymptotic) mass of the system and
the
physics of the mixture of gas and radiation. Then it would be natural to
ask:
what mass is within the compact body?
Alternatively, the mass of the core would be known and the total mass
would
require determination.

The main goal of this paper is the numerical investigation of this problem
for stationary flows. We assume spherical symmetry and adopt thin gas
approximation
in the transport equation. It is already known from studies of newtonian
radiation hydrodynamics \cite{AA} --- \cite{Roszkowski} that supersonic
flows
are generically not fixed by total luminosity, asymptotic temperature and
redshift. To each set of such data there can correspond two solutions with
different gas abundances. Changing luminosity one obtains two curves, on
the
luminosity-(gas abundance) diagram, that originate at a bifurcation point.
This
point is unique, for given boundary data. General-relativistic supersonic
flows
with small redshifts are similar to newtonian ones in that they also
branch from
a bifurcation transonic flow. In the case of high redshifts supersonic
general-relativistic flows can be absent. A similar picture appears in
transonic flows of perfect gases, newtonian or general-relativistic,
without
radiation. In this case boundary data can consist of the mass accretion
rate and the asymptotic speed of sound \cite{PRD2006} and the only unique
solution --- a branching point --- corresponds to the maximal accretion.

Accretion systems with subsonic flows are not determined by the data
described hitherto. One needs additional information, for instance the asymptotic
gas density, in order the specify the solution completely. We discover,
however, an interesting fact valid in the newtonian case and in the low-redshift
regime of general relativity: transonic flows encompass, on the luminosity-(gas
abundance) diagram, the set filled with subsonic flows. Therefore the two
transonic branches provide estimates of the mass abundance of
corresponding
subsonic solutions. In particular, numerical analysis suggests that the
most
luminous flow is supersonic. This picture is valid in the newtonian level
and
also in the general-relativistic case, for small redshifts. If redshifts
are
large, then the boundary of the set of subsonic solutions may consist of
transonicor subsonic flows, but it is remarkable that the shape of the
set of
subsonic solutions is only weakly dependent on redshifts. In particular,
the
flow with maximal luminosity is unique.

This investigation can be useful in the analysis of two important issues.
There
is the question of identification of the so-called Thorne-\.Zytkow stars
\cite{Thorne}, that consist of a hard core and overblown atmosphere. They
are
conjectured to result in the merger of a main sequence star with a neutron
star
or a black hole. If the core consists of a neutron star, then its mass is
roughly known. Results of this paper show that one can estimate the total mass
 by measuring luminosity, asymptotic temperature and redshifts.
Another interesting application would be to distinguish compact stars
(neutron
stars or gravastars \cite{Mazur}) from black holes, but the present
analysis
would require further elaboration. Within the scenario investigated here
it
does not seem feasible to distinguish between a black hole or a gravastar,
but
the investigation of stability can possibly give further information.

The organization of the rest of this paper is following. Section 2
presents
spherically symmetric equations of radiation hydrodynamics. The next
section
explains the concept of quasistationary solutions. The interaction of gas
and
radiation is treated in the thin gas approximation \cite{Mihalas}. The
final
form of required equations is displayed. Boundary conditions are described
in
Section 4. The next Section brings a discussion of boundary conditions. In
particular, we explain the relation between the binding energy of
collapsing
fluids and radiation redshifts. We demonstrate in Section 6 that
supersonic
solutions constitute a one-sided boundary for the set of subsonic
solutions in
newtonian test fluids. Section 7 shows main results of this paper ---
relations
between transonic and subsonic flows in the general-relativistic case.
The
last section contains a summary.

\section{Equations of general-relativistic hydrodynamics.}
\label{sec:2}
We use comoving coordinates $t, r, 0\le \theta \le \pi , 0\le \phi <
2\pi$:
time, coordinate radius and two angle variables, respectively. The metric
can
be chosen in the form
\begin{equation}
ds^2=-N^2dt^2+\tilde adr^2 +R^2d\Omega^2.
\label{1}
\end{equation}
$R$ in (\ref{1}) is the area radius. The infall velocity of gas is equal
to
$U = \frac{1}{N} \frac{dR}{dt}$ and it is related to extrinsic curvatures
of
the Cauchy hypersurface $t=\const$,
\begin{equation}
\left(\trK - K^r_r\right) R= 2U.
\label{2}
\end{equation}
$\trK$ is the trace of the extrinsic curvature and $K_r^r$ is its
radial-radial
component.

The energy-momentum tensor reads $T_{\mu \nu }=T_{\mu \nu }^B +T_{\mu \nu
}^E$,
where the baryonic part is given by $T_{\mu \nu }^B =
(\rho +p )U_\mu U_\nu +pg_{\mu \nu }$ with the time-like and normalized
four-velocity $U_\mu $, $U_\mu U^\mu =-1$.
The radiation part has only four nonzero components, $T_0^{0E}\equiv
-\rho_E=-T_r^{rE}$ and $T^E_{r0}= T^E_{0r}$, which is consistent with
 the so-called thin gas approximation (see next section).

A comoving observer would measure local mass densities, material
$\rho =T^{B\mu \nu }U_\mu U_\nu $ and radiation $\rho_E$, respectively.
The
baryonic current is defined as $j^\mu \equiv \rho_0 U^\mu$, where $\rho_0$
is
the baryonic mass density. Define $n_\mu$ as the unit normal to a centered
(coordinate) sphere lying in the hypersurface $t=\const$ and $k$ as the
related
mean curvature scalar, $k={R\over 2}\nabla_i
n^i=\frac{1}{\sqrt{\tilde a}}\partial_rR$.
The comoving radiation flux density reads
$j = U_\mu n^\nu NT^{\mu E}_\nu /\sqrt{\tilde a} = NT^{0E}_r /\sqrt{\tilde a}$.

We assume a polytropic equation of state for the baryonic matter,
$p=K\rho_0^\Gamma $ ($K$ and $\Gamma $ are constants) and the relation
\begin{equation}
\rho =\rho_0+h,
\label{3a}
\end{equation}
where the internal energy $h$ is easily shown to be equal to $p/(\Gamma -1)$.
The equations of motion consist, in the spherically symmetric case, of
three
Einstein equations, of two constraints \cite{Iriondo}
\begin{equation}
\frac{1}{R} \partial_R \left( R k^2 \right) =
- R \Bigl(8\pi \left( \rho +\rho_E\right) 
+ \frac{3}{4} (K_r^r)^2 \Bigr) + \frac{1}{R} +
\frac{R}{4}(\trK)^2 + \frac{R}{2} \trK K_r^r,
\label{3}
\end{equation}
\begin{equation}
\frac{\partial_r (K_r^r - \trK)}{\sqrt{\tilde a}} = -\frac{3}{R} k K_r^r - 8\pi j
+
\frac{1}{R} k \trK,
\label{5}
\end{equation}
and one dynamical equation
\begin{eqnarray}
\lefteqn{\partial_t( K_r^r - \trK )= \frac{3N}{4}(K_r^r)^2
- \frac{N k^2}{R^2} - \frac{2k}{R\sqrt{\tilde a}} \partial_r N +
\frac{N}{R^2} }\nonumber\\
&& +8\pi N T_r^r + \frac{3}{4} N (\trK)^2 -
\frac{3N}{2} \trK K_r^r.
\label{6}
\end{eqnarray}

The baryonic current is conserved,
\begin{equation}
\nabla_\mu j^\mu = 0.
\label{7}
\end{equation}
There are four conservation equations $\nabla_\mu T^{\mu B}_\nu =
-\nabla_\mu T^{\mu E}_\nu ={\mathcal F}_\nu$ (here $\nu = 0, r$). The quantity 
${\mathcal F}_\nu $ is called the radiation force density and it describes interaction between
baryons and radiation. Its radial component will be written as
 ${\mathcal F}_r\equiv k\sqrt{\tilde a }F_r$; $F_r$ is defined later.
The formulation of general-relativistic radiation hydrodynamics
presented here agrees with that of Park \cite{Park} and Miller and Rezzola
\cite{Rezzola}, and (on a Schwarzschildean background) with Thorne
\textsl{et.\ al} \cite{ThorneF}. One can solve formally both constraints,
arriving at \cite{Iriondo}
\begin{eqnarray}
&& k = \sqrt{1-\frac{2m(R)}{R} +U^2}, \nonumber\\
&& K_r^r = \partial_RU - 4\pi R \frac{j}{k}, \nonumber\\
&& \trK = {1\over R^2}\partial_R\left( U R^2 \right) - 4\pi R \frac{j}{k}.
\label{8}
\end{eqnarray}
Here $m(R)$ is the quasilocal mass
\begin{equation}
m(R)=m (R_\infty )-4\pi \int_R^{R_\infty }dr r^2\left( \rho +\rho_E +
\frac{Uj}{k}\right) .
\label{9}
\end{equation}
The integration in (\ref{9}) extends from $R$ to the outer boundary
$R_\infty$
of the accretion system. $m(R_\infty )$ can be equal (or arbitrarily
close) to
the total asymptotic mass $M$. The contribution to the mass coming from
the
exterior of $R_\infty$ can be neglected. Comoving coordinates can be
understood
as a choice of a particular integral-type gauge condition \cite{EM99}. In
what
follows we will use the comoving spacetime foliations, with the time $t$
but
often with the areal radius $R$ instead of the comoving radius $r$. The
parametrization ($t$, $R$) can be interpreted as corresponding to an
observer at
rest at $R$.

One can choose an alternative set of coordinates in a different spacetime
foliation, in the so-called polar gauge $\trK = K_r^r$ (no summation),
with the
time $t_S = t_S(R,t)$ and the areal variable $R$. We do not employ these
variables
here (but see a remark below).

A simple but lengthy calculation shows that the local mass changes
according to
the following rule
\begin{equation}
\partial_{t_S} m(R)= \left( \partial_t -NU\partial_R \right) m(R)=
4 \pi \left( Nk R^2\left( j\left( 1+ \left( \frac{U}{k}\right)^2\right)
+ 2\rho_E{U\over k} \right) +
NU R^2 \left( \rho +p \right) \right)_{R}^{R_\infty} +A_\infty ,
\label{10}
\end{equation}
where $A_\infty $ is the value of $-4\pi NU R^2\left( \rho +\rho_E +
\frac{Uj}{k}\right) $ at $R=R_\infty $.
It is interesting to note that the expression
$4\pi Nk R^2\left( j\left( 1+ \left( \frac{U}{k}\right)^2\right)
+2U\rho_E/k
\right)$ represents the radiation flux measured by an observer located at
$R$
in the polar gauge foliation. The mass
contained in the annulus $(R, R_\infty )$ changes if the fluxes on the
right hand side  do not cancel. In the case of quasistationary flows
the mass is approximately constant.

\section{\label{sec:3}Quasistationary flows in the thin gas
approximation.}
We will say that the accretion process is quasistationary if all relevant
quantities measured in the rest frame are approximately constant during
time
intervals much smaller than certain characteristic time scale $T$. In
analytical terms, we assume that $\partial_{t_S}X= (\partial_t -
NU\partial_R) X = 0$ for $X=\rho_0$, $\rho$, $j$, $U$\ldots

Under quasistationarity assumption the evolution equation (\ref{6}) can be
written as
\begin{equation}
U \frac{dU}{dR} = \frac{k^2}{N} \frac{dN}{dR} - \frac{m(R)}{R^2} -
4\pi R \left( p+\rho_E \right)
\label{12}
\end{equation}
From this one easily obtains
\begin{equation}
N = k \exp \left( -4\pi \int_R^{R_\infty } dr \frac{r}{k^2}
\left( \rho + p +2\rho_E + \frac{Uj}{k} \right) \right) .
\label{13}
\end{equation}
The local accretion mass rate $\Mdot $ is defined as 
\begin{equation}
\Mdot \equiv - 4\pi U R^2 \rho_0.
\label{11}
\end{equation}
The baryonic current conservation equation (\ref{7}) takes the form
$\partial_R \Mdot = -16\pi^2 j R^3 \frac{\rho_0}{k}$; thus the local mass
accretion rate $\Mdot $ can change, but the quantity
\begin{equation}
\tilde M \equiv - 4\pi U R^2 \rho_0 -(4\pi )^2 \int_R^{R_\infty } dr r^3 j
\frac{\rho_0}{k}
\label{14}
\end{equation}
remains constant, $\frac{d}{dR} \tilde M=0$. Notice that $\Mdot \left(
R_\infty \right) = \tilde M$.

As the characteristic time scale one can choose the quantity related to
the runaway instability, $T\equiv M/\tilde M$.  
One can obtain a rough estimate $\frac{d}{dt}\tilde M\le C M^2$ where $C$
is a
constant (see \cite{AA} for the corresponding derivation in the newtonian
case).
Let $M_0$ be the initial mass. Then $M\le 1/\left( 1-CM_0t\right)$. If
$M/\tilde M \gg t$ then $M\approx M_0$ and if $t\approx T$ then $M\gg
M_0$;
thus the time $T$ sets the time scale for the runaway instability.

The radiation force density has only one nonzero component. It is a simple
exercise to show that the conservation of $\tilde M$ and relation
(\ref{3a})
imply, for polytropic gases, the vanishing of the zeroth component of the
radiation force, ${\mathcal F}_0$. Therefore the two
related
energy and radiation energy balance equations read
\begin{eqnarray}
N U \frac{d \rho}{dR} + N \trK \left( p+\rho \right) & = & 0 \nonumber\\
\frac{N}{UR^2} \frac{d}{dR} \left( U^2R^2\rho_E\right) + \frac{k}{R^2
N}
\frac{d}{dR} \left( N^2R^2j \right) - 8\pi N Rj \frac{\rho_E}{k} & = & 0.
\label{15}
\end{eqnarray}
The two other energy-momentum conservation equations are displayed below.
The
relativistic Euler equation is given by
\begin{equation}
\frac{1}{N} \frac{dN}{dR} \left( p+\rho \right) + \frac{dp}{dR} = F_r
\label{16}
\end{equation}
and the transport equation reads
\begin{equation}
\frac{N}{U R^2} \frac{d}{dR} \left( U^2 R^2 j\right) + \frac{k}{R^2 N}
\frac{d}{dR} \left( N^2 R^2 \rho_E \right) - 8\pi R j^2 \frac{N}{k} =
-NkF_r.
\label{17}
\end{equation}
There enters an important phenomenological assumption that can be easily
expressed in terms of quantities related with comoving coordinates. Namely
we assume the so-called thin gas approximation \cite{Mihalas}
\begin{equation}
F_r=\kappa  \rho_0 j.
\label{18}
\end{equation}
The only direct interaction between baryons and radiation is through
elastic
Thompson scattering. $\kappa $ is a material constant, depending in
particular
on the Thompson cross section $\sigma $, $\kappa =\sigma /\left( 4\pi m_p
c
\right)$. $c$ is the speed of light and $m_p$ is the proton mass.

In summary, the complete set of equations would consist of eqs (\ref{8}),
(\ref{9}) and (\ref{13}--\ref{18}).

\section{Final equations.}
It is convenient to express all relevant quantities in terms of the speed of
sound $a$, given by $a^2\equiv {dp\over d\rho }$, because we will search for flows possessing 
transonic points. We find that it is computationally expedient to replace the radiation
energy balance equation (the second of eqs (\ref{15})) by the total energy
conservation
\begin{equation}
\Mdot N \frac{\Gamma-1}{\Gamma-1-a^2}+ 2 \Mdot N \frac{\rho_E}{\rho_0} =
4\pi R^2 j N k \left( 1+ \frac{U^2}{k^2}\right) +C .
\label{19}
\end{equation}
Equation (\ref{19}) is a direct consequence of quasistationarity and the
equation (\ref{10}). The constant $C$ is the asymptotic flux
(\textsl{i.e.},
flowing through the sphere of a radius $R_\infty $) in (\ref{10}).

Furthermore, write down eq.\ (\ref{17}) as
\begin{eqnarray}
\lefteqn{\frac{NU}{R^2} \frac{d}{dR} \left( R^2 j\right) =-\kappa kN j \rho_0-
\frac{kN}{R^2} \frac{d}{dR} \left( R^2\rho_E\right) - }\nonumber\\
&& 2Nj \frac{dU}{dR} - 2k \rho_E \frac{dN}{dR} + 8 \pi NR \frac{j^2}{k}.
\label{20}
\end{eqnarray}
and replace the term $\frac{NU}{R^2} \frac{d}{dR} \left( R^2j\right)$ in
the
second of equations (\ref{15}). After some algebra one arrives at
\begin{eqnarray}
\lefteqn{\left( 1- \frac{2m(R)}{R} \right) \frac{N}{R^2} \frac{d}{dR}
\left( R^2\rho_E\right)=-\kappa k^2N j\rho_0 + 
2N\left( U \rho_E -kj\right) \frac{dU}{dR} +} \nonumber\\
&& 2k\left( jU -k\rho_E\right) \frac{dN}{dR} + 8\pi NR
\left( j^2 - j \rho_E \frac{U}{k}\right) .
\label{21}
\end{eqnarray}
  Below we display a number of relations, that can
be easily obtained from the equation of state and (\ref{3a}), namely
\begin{eqnarray}
\frac{p}{\rho_0} & = & \frac{\Gamma-1}{\Gamma } \frac{a^2}{\Gamma-1-a^2},
\nonumber\\
\frac{p + \rho}{\rho_0 } & = & \frac{\Gamma-1}{\Gamma-1-a^2}, \nonumber\\
\frac{\partial_R p}{p+\rho } & = & \partial_R \ln
\left( \frac{\rho+p}{\rho_0} \right)
= - \partial_R \ln \left( \Gamma-1 -a^2\right) , \nonumber\\
\rho_0 & = & \rho_{0\infty }
\left( \frac{a}{a_\infty }\right)^\frac{2}{\Gamma-1}
{\left( 1 - \frac{a^2_\infty}{\Gamma -1}\right)^\frac{1}{\Gamma-1}\over
\left( 1 - \frac{a^2}{\Gamma -1}\right)^\frac{1}{\Gamma-1}}.
\label{22}
\end{eqnarray}
It is useful to insert (\ref{22}) into Eqs.\ (\ref{14}---\ref{18}). One
obtains
following equations:
\begin{enumerate}
\renewcommand{\theenumi}{\roman{enumi}}
\renewcommand{\labelenumi}{\theenumi )}
\item the gas energy density conservation equation (the first of eqs.\ (\ref{15}))
\begin{eqnarray}
\lefteqn{ \frac{d}{dR} \ln a^2 = -\frac{\Gamma-1-a^2}{a^2- \frac{U^2}{k^2}}
\times }\nonumber\\
&&\Biggl( \frac{1}{k^2 R} \left( \frac{m}{R} -2U^2 + 4\pi R^2
\left( \rho_E + p + j \frac{U}{k} \right) \right) - \nonumber\\
&&\kappa j \left( 1- \frac{a^2}{\Gamma-1} \right) \Biggr) ;
\label{23}
\end{eqnarray}
\item the baryonic mass conservation
\begin{equation}
\frac{dU}{dR} = -\frac{U}{\Gamma -1 -a^2} \frac{d}{dR} \ln a^2 -
\frac{2U}{R} +
\frac{4\pi R j}{k};
\label{24}
\end{equation}
\item the equation for the lapse
\begin{equation}
\frac{dN}{dR}=N \left( \kappa j \frac{\Gamma-1-a^2}{\Gamma-1} +
\frac{d}{dR}
\ln \left( \Gamma -1 -a^2 \right) \right) .
\label{25}
\end{equation}
Eq.\ (\ref{25}) follows from (\ref{16}) and (\ref{22}).
\end{enumerate}

Equations (\ref{19}), (\ref{21}) and (\ref{23}---\ref{25}) constitute,
with $k$
and $m(R)$ given by (\ref{8}) and (\ref{9}), the complete model used in
numerical calculations reported in next sections. A remark is in order. It
is
clear from the inspection of (\ref{23}) that if $a^2=\frac{U^2}{k^2}$ (the
speed of sound equals the spatial length of the infall velocity) then the
expression $\frac{1}{k^2 R} \left( \frac{m}{R} - 2U^2 + 4\pi R^2
\left( \rho_E + p + j \frac{U}{k} \right) \right) -\kappa j \left(
1 - \frac{a^2}{\Gamma-1} \right)$ must vanish. There are four different
ways of
passing through the transonic point (similarly as in the newtonian
analysis in
\cite{Bondi}) and only one of those corresponds to the accretion.

\section{Boundary conditions.}
The overall picture of the system is as follows. A ball of gas is enclosed
by a
sphere $S_\infty$ of a radius $R_\infty$ and connected, via a narrow
transient
zone filled with baryonic matter and radiation, to the Schwarzschild vacuum
spacetime. It is clear that by careful arrangement of data the mass within
the
transient zone can be negligible. Therefore we assume that the asymptotic
mass
$M$ is equal to $m\left( R_\infty \right)$ (see Eq.\ (\ref{9})).

Boundary conditions at the outer sphere $S_\infty $ are needed for the
radiation quantities $j$, $\rho_E$, the mass accretion rate $\Mdot$, the
square
of the speed of sound $a^2_\infty$ and the baryonic mass $\rho_\infty$. We
assume that $a^2_\infty \gg M/R_\infty \gg U^2_\infty$; the second
inequality
means that infall velocity is much smaller than the escape velocity. These
inequalities guarantee the fulfillment of the Jeans criterion for the
stability
(see a discussion in \cite{AA} and studies of stability of accreting flows
in
newtonian hydrodynamics \cite{Mach}), suggesting the stability of
solutions.

One can derive, after some algebra involving manipulations of equations
(\ref{20}) and (\ref{21}), the approximate equation $\frac{d}{dR} \left(
\left(
\rho_E -j\right) R^2\right) \approx 0$ in the asymptotic region. This
means,
taking into account the fact that $R_\infty$ can be arbitrarily large,
that one
can safely assume $j_\infty =\rho_{E_\infty}$. Furthermore, the total
luminosity
is with good accuracy given by $L_0=4\pi R^2_\infty j_\infty $ and it must be
related
to the accretion rate by the formula
\begin{equation}
L_0\equiv \alpha \Mdot_\infty  .
\label{26}
\end{equation}
The coefficient $\alpha$ determines the relative binding energy. We assume
\begin{equation}
\alpha \equiv 1- \frac{N\left( R_0\right)}{k\left( R_0\right)}
\sqrt{1 - \frac{2m\left( R_0\right)}{R_0}},
\label{27}
\end{equation}
where $R_0$ is the outer radius of the hard core of the system. The last
two
formulae can be justified by two arguments. First, in the nonrelativistic
limit
one gets $\alpha =|\phi \left( R_0\right) |$; $\alpha$ is equal to the
absolute
value of the newtonian potential on the surface of the hard body. It is
clear
now that (\ref{26}) is just the statement that all available binding
energy is
transformed into radiation, and that there is an implicit assumption that
the
heat capacity of the core is negligible. Second, the condition of
stationarity
implies the existence of the approximate time-like Killing vector. By
employing
standard reasoning \cite{Schutz}, one arrives at the two formulae
(\ref{26})
and (\ref{27}). Thus $\alpha$ can be regarded as a proper binding energy.
Let
us remark that $\alpha$ gives the standard measure of the gravitational
red- or
blue-shift. If stationary observers detect $\omega_0$ at $R_0$ and
$\omega$ at
infinity, and $1/\omega \ll 2M$ (the geometric optics condition --- see
\cite{Karkowski2003} for a discussion) then $\omega = \left( 1-\alpha
\right)
\omega_0$.

In conclusion, boundary data consist of the binding energy coefficient
$\alpha$,
total luminosity $L_0$ and the asymptotic speed of sound $a^2_\infty$. Not
only
$L_0$ but also the two remaining quantities can be in principle determined
from
observations: $\alpha$ from the measurement of the highest redshift and
$a^2_\infty$ from the asymptotic temperature. Then $j_\infty =
\rho_{E_\infty} =
L_0/\left( 4\pi R^2_\infty \right)$, and the mass accretion rate
$\Mdot = L_0/\alpha$. These data in fact specify transonic flows up to,
possibly, a bifurcation; for given data there can exist two solutions. In
the
case of subsonic flows another boundary condition is needed, for instance
the
baryonic mass density $\rho_\infty$. We show later that transonic flows can give
bounds onto some characteristics of relevant subsonic solutions.

\section{Subcritical versus critical: lessons from newtonian hydrodynamics
of
test fluids.}

In the Bondi model \cite{Bondi} the selfgravity of gases is neglected. The
relevant equations can be obtained from these presented above in the
following
way: assume the test gas approximation, $\alpha = j=\rho_E=0$, $k=N=1$ and
$\Gamma-1-a^2\approx \Gamma-1$. In this approximation $\rho =\rho_0$. The
whole
problem reduces to two algebraic equations
\begin{equation}
\dot M =C
\label{28}
\end{equation}
\begin{equation}
\frac{U^2}{2} + \frac{a^2}{\Gamma -1} - \frac{M}{R} =
\frac{a^2_\infty}{\Gamma -1}.
\label{29}
\end{equation}
From (\ref{28}) one has $U=\frac{-\Mdot}{4\pi \rho R^2}$. Since now
$\rho = \rho_\infty \left( {a^2\over a^2_\infty }\right)^\frac{1}{\Gamma
-1}$,
one obtains
\begin{equation}
U^2 =2\frac{\beta}{R^4}\left(
\frac{a^2_\infty}{a^2}\right)^\frac{2}{\Gamma-1},
\label{30}
\end{equation}
where $\beta \equiv \frac{\Mdot^2}{32\pi^2 \rho_\infty^2}$. Insertion of
(\ref{30}) into (\ref{29}) yields
\begin{equation}
\frac{\beta}{R^4}\left( \frac{a^2_\infty}{a^2} \right)^\frac{2}{\Gamma-1}
+
\frac{a^2}{\Gamma-1} - \frac{M}{R} = \frac{a^2_\infty}{\Gamma-1}.
\label{31}
\end{equation}
Thus newtonian transonic flows have this interesting property that the
quantity $\beta$ is given by a simple analytic formula involving boundary
data:
\begin{equation}
\beta \equiv \beta_c ={ M^4\over 32 a^6_\infty } \left( {2 \over 5-3\Gamma
}
\right)^{5-3\Gamma \over \left( \Gamma -1\right) };
\label{31a}
\end{equation}
that means that, given $\Mdot$ and $a_\infty$, the density $\rho_{\infty
}$
and $\rho $ is also specified. One finds from Eq.\ (\ref{31}) that
\begin{equation}
\frac{d a^2}{d\beta} = -\left( \Gamma -1\right) \frac{U^2 a^2}{\beta }
\frac{1}{a^2-U^2}
\label{32}
\end{equation}
and there exists (from the implicit function theorem \cite{Schwarz}) a
local
solution $a^2(\beta)$. One infers from (\ref{32}) that outside the
supersonic
sphere the speed of sound decreases with the increase of $\beta$. Taking
that
into account, one concludes from the inspection of (\ref{30}), that the
infall
velocity increases with the increase of $\beta$. Therefore subsonic flows,
for
which $U^2< a^2$, can exist only for values of the parameter $\beta$
smaller
than $\beta_c$. That means that, given a transonic and subsonic flows
with the
same data $a_\infty$, $\Mdot$, the asymptotic mass density of the subsonic
flow
must be bigger than that ($\rho_{c \infty }$) of the (test fluid) transonic flow,
\begin{equation}
\rho_{\mathrm{sub} \infty} > \rho_{\mathrm{c} \infty }.
\label{33}
\end{equation}
Notice that the asymptotic mass density of transonic flows can be
represented
as $\rho_{\mathrm{c}\infty } \equiv \beta_\mathrm{c}/\Mdot$; therefore the
mass
density is now completely specified by the other data. Another way of
rephrasing (\ref{33}) is to say that test fluid    flows  are more efficient than
subsonic ones in the sense that a given mass accretion rate demands less
gas in
the first case than in the other. See also another derivation in
\cite{Padmanabhan}. One can infer from this description that the branch of
test fluid transonic flows embraces from below the set of subsonic solutions,
in
the (mass accretion rate)-(gas abundance) diagram.

The same conclusion holds true for the Shakura model \cite{Shakura},
assuming
the test fluid approximation. Indeed, in this case one has instead of
Eq.\ (\ref{31a}) the following \cite{AA}
\begin{eqnarray}
\lefteqn{ |\phi (R_0)|\left( \exp \left( -{GL_0M\over R|\phi (R_0)|L_E }
\right) -1\right) =} \nonumber \\
&& \frac{\beta}{R^4}
\left( \frac{a^2_\infty}{a^2 }\right)^\frac{2}{\Gamma-1} +
\frac{a^2}{\Gamma-1} -\frac{M}{R} - \frac{a^2_\infty}{\Gamma-1},
\label{34}
\end{eqnarray}
where only $a^2$ and $\beta$ depend on the mass parameter $\rho_{\infty
}$.
Here $\phi (R_0)$ is the newtonian potential on the surface of the compact
core
and $G$ denotes the gravitational constant. One can find the dependence of
$a^2$
as the function of $\rho_\infty$ in a similar way as before. Thus a
supersonic
flow with a given luminosity can have less gas than a subsonic
flow of the same luminosity.

In particular, again we observe that subsonic flows lie above the
supersonic
(test fluid) branch, in the luminosity-(gas abundance) diagram. Numerical
studies show more, that the set of subsonic solutions has a
parabola-shaped
boundary that consists only of transonic solutions. The edge of the
parabola
has maximal luminosity.

\section{Numerical results.}

Below we will study how much of the newtonian picture drawn in the
preceding
section is valid in the general-relativistic case.

The equations are put in the evolution form, see eqs.\ (\ref{19}),
(\ref{21}),
(\ref{23}-\ref{25}). We start from the outer boundary $R_\infty$ and evolve
inwards until the equality
\begin{equation}
\alpha = 1- \frac{N\left( R\right)}{k\left( R\right) }
\sqrt{1-\frac{2m\left( R\right)}{R}}
\end{equation}
is met. The corresponding value of the radius is denoted as $R_0$ and it
is
regarded as the size of the inner core. It appears that the 4th order
Runge-Kutta method fails almost immediately and for that reason we
employed the 8th order Runge-Kutta method \cite{RK8}.

We assume required boundary data
$j_\infty =\rho_{E_\infty} $, the mass accretion rate $\Mdot_\infty $, the speed of sound
$a^2_\infty$, the
parameter $\alpha $ and the baryonic mass density $\rho_\infty$.
Inequalities  $U^2_\infty \ll M/ R_\infty \ll a^2_\infty$ are obeyed,
since this ensures the  Jeans length of the configuration to be  much larger
than $R_\infty $, which in turn suggests stability of the configuration (see a discussion in
\cite{AA}). Solutions are obtained by the method of shooting. Given $a^2_\infty$, $\alpha$ and
$L_0=\alpha \Mdot_\infty $, one should vary the asymptotic baryonic mass density
$\rho_{0 \infty}$.   We find that solutions  exist  for a finite range of $b_1\le \rho_{0 \infty}
\le b_2$. The change of the luminosity $L_0$, while keeping constant
$a^2_\infty$ and  $\alpha$,  results in another segment of solutions with baryonic 
densities $b_1'\le \rho_{0 \infty}\le b_2'$. Solutions corresponding to the extremal
baryonic densities $b_1(L_0), b_2(L_0)$ are shown in Figs 1-3 as the bifurcation curves.
 Each point on the first three figures  (1-3)
within the region enclosed by bifurcation curves represents a solution with a specific 
value of the asymptotic baryonic density and fixed  boundary data ($a^2_\infty$, $\alpha$ and
$L_0=\alpha \Mdot_\infty $).    

The numerical integration is straightforward with the important exception
of  transonic solutions. It is clear from Eq.\
(\ref{23}) that the flow becomes critical at a sonic point and if
$a^2=U^2/k^2$
then
\begin{equation}
\frac{1}{k^2 R} \left( \frac{m}{R} -2 U^2 + 4\pi R^2 \left( \rho_E + p +
j \frac{U}{k} \right) \right) = \kappa j \left( 1-
\frac{a^2}{\Gamma-1}\right).
\label{35}
\end{equation}
The numerical strategy for finding transonic flows is as follows.
For a density $\rho_{0\infty } $ chosen at random one either obtains no solution at all or a
subsonic solution. Using the bisection method one can obtain a boundary of the
solution set, later on called the bifurcation curve. This search process can be
automated and it works well for all values of the parameter $\alpha$. For small
parameters $\alpha$ one gets convincing numerical evidence that the bifurcation
curve consists solely of transonic solutions.

When $\alpha$ is close to one, then the automated search produces a
bifurcation curve,
but the question whether it consists of transonic flows has to be studied
in a more detailed way.
If $\alpha$ is significant then the numerical problem becomes quite
sensitive
on tiny deviations -- of the order of $10^{-15}\rho_0$ -- from the right
values
of the asymptotic mass density. The bifurcation curve is found within some
margin error and that error would in many cases be larger than
$10^{-15}\rho_0$;
thence automated search becomes inconclusive. Investigating collected data
one can in some cases determine the character of a solution that lies on the
bifurcation curve.

Another subtle problem is evolving Eq.\ (\ref{23}) in the vicinity of the sonic
point. The analytic reason is due to the fact that the sonic point is a
critical spatial point at which coalesce four different solutions: two
accretion branches and two wind branches, in each case one inside and one
outside of the sonic sphere. This agrees with the well known feature of
the standard Bondi accretion of test fluids \cite{Bondi}. The accretion flow
solution consists of two branches, that existing outside of the sonic
sphere and the other that bifurcates inward from the sonic point. The accretion
branch is unstable beneath the sonic sphere. From the numerical point of view
when the
denominator $a^2-U^2/k^2$ of Eq.\ (\ref{23}) is small (smaller than
$10^{-12}$)
then the whole fraction is calculated with a large error. For that reason
there
must exist a procedure for checking the value of $a^2-U^2/k^2$ at a point
and
if it becomes too small then a regularization method must be implemented.
It
appears that it is enough to do this regularization only during one step
while passing through the sonic point.

We assumed following asymptotic parameters.  We choose $M_\odot /M=
5.95496\times 10^{-7}$, where $M_\odot $ is the Solar mass.  Let us remark here 
that, while a definite  choice is needed by the nature of numerical
calculation, there is nothing pecular in the above data. One can repeat 
this analysis assuming that the total mass is of the order of the Solar mass.

The parameter $\kappa = \sigma /\left( 4\pi m_p c\right) $, as we explained earlier.
In this paper we adopt the traditional choice of units
$G=c=1$ and supplement it  by the scaling $M=1$. This leads to the value
of the
$\kappa = 2.1326762 \times 10^{21} \left( M_\odot /M\right)$, that is
$\kappa =1.27\times 10^{15}$. The size of the system is $R_\infty =10^6$.
The speed of sound is given by $a^2_\infty =4\times 10^{-4}$ in all numerical
calculations that are described below. The Eddington luminosity reads
$L_E=4\pi M/\kappa = 9.9847 \times 10^{-15}$.

\begin{figure}[h]
\includegraphics[width=\linewidth]{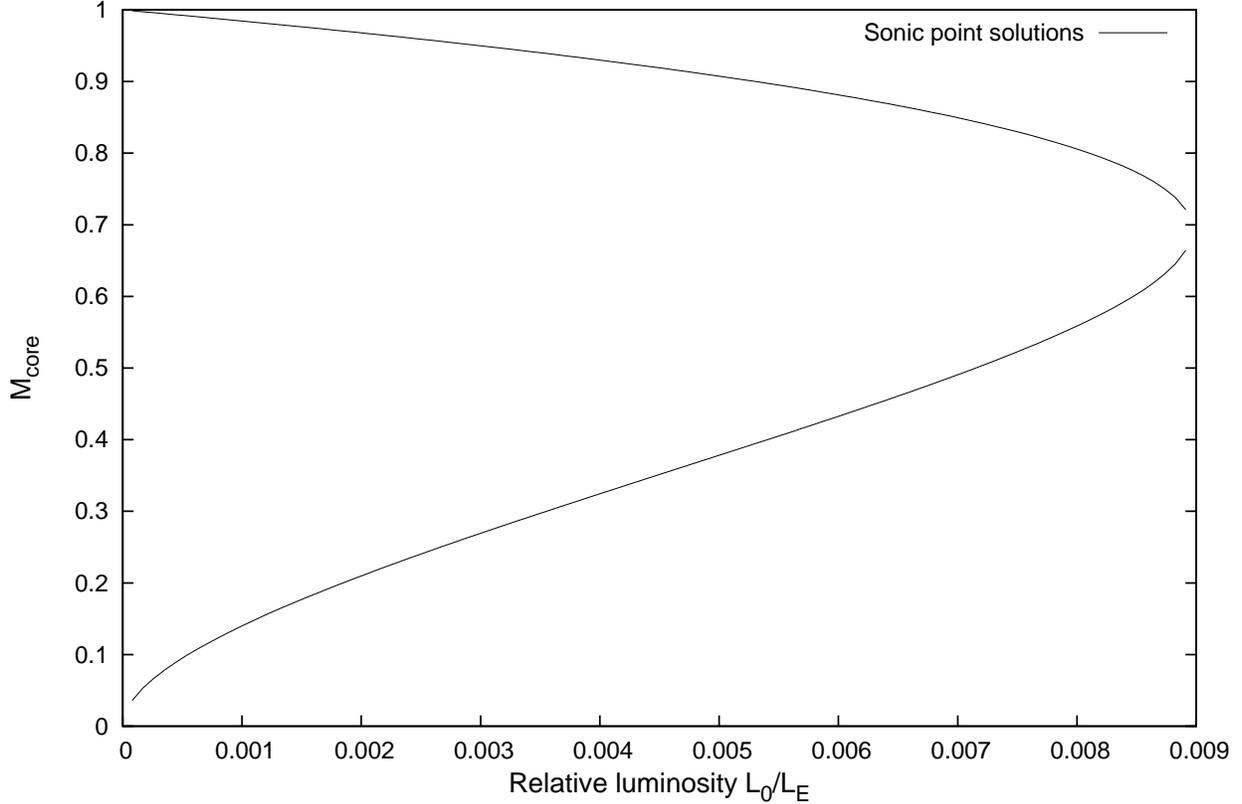}
\caption{\label{fig:1}Bifurcation curve for  small binding energy,
$\alpha = 0.0025$. Two
branches of
transonic flows encompass the set of subsonic flows. The abscissa shows
the
luminosity and the ordinate shows the mass of the compact core.}
\end{figure}

\begin{figure}[h]
\includegraphics[width=\linewidth ]{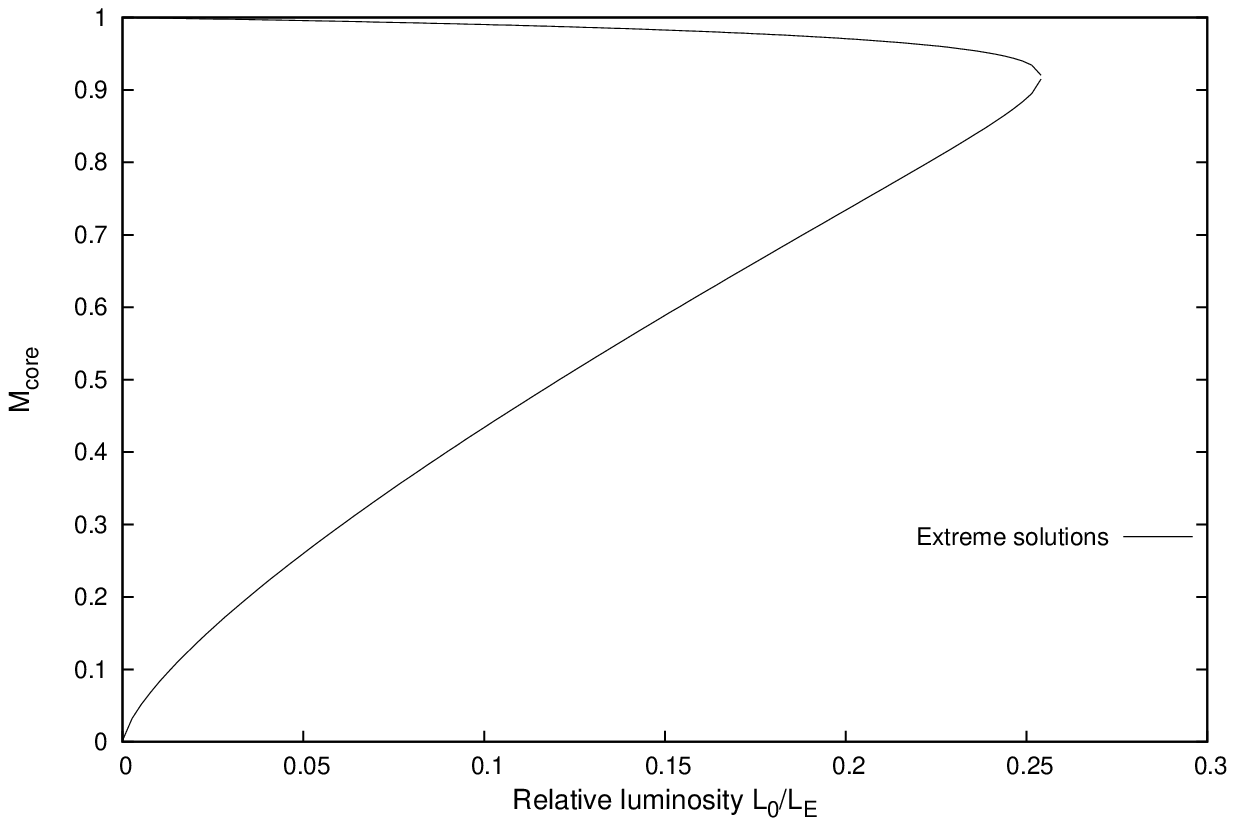}
\caption{\label{fig:2}Bifurcation curve for intermediate binding energy, $\alpha = 0.5$. Two
bifurcation branches
encompass the set of subsonic flows. The abscissa shows the
luminosity and the ordinate shows the mass of the compact core.}
\end{figure}

\begin{figure}[h]
\includegraphics[width=\linewidth]{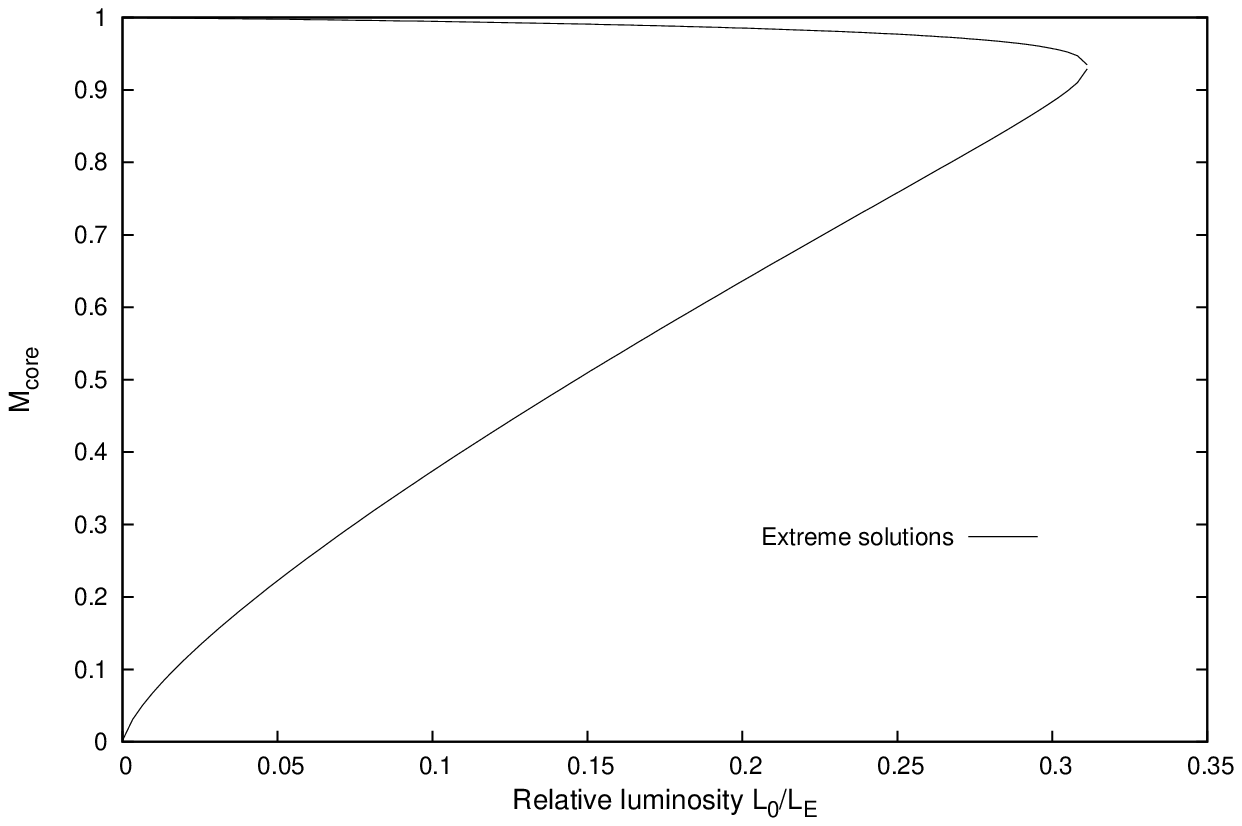}
\caption{\label{fig:3}Bifurcation curve for high binding energy, $\alpha = 0.9$. Two bifurcation branches
 encompass the set of subsonic flows. Abscissa shows the
luminosity and the ordinate shows the mass of the compact core.}
\end{figure}

Figures \ref{fig:1}--\ref{fig:3} show accreting solutions on the diagram
luminosity-(mass of the central core). Each point within the set embraced 
by two curves (bifurcation curves) corresponds to an accretion solution.
Solutions are absent outside of this region.   Typically, for small binding energies
(that is, small $\alpha$) and small luminosity $L_0$, there exist two accreting
solutions possessing sonic points, with asymptotic densities $\rho_{0 \infty 1}$
and $\rho_{0 \infty 2}$. There appear subsonic flows for each $\rho_{0\infty }
\in \left(\rho_{0 \infty 1 }, \rho_{0 \infty 2 } \right)$. Thus the two
bifurcation branches of transonic flows embrace a set of subsonic flows.
Accreting stationary flows are absent above the bifurcation point, which
maximizes the luminosity. All that is illustrated in Figure \ref{fig:1}.

\begin{figure}[h]
\includegraphics[width=\linewidth]{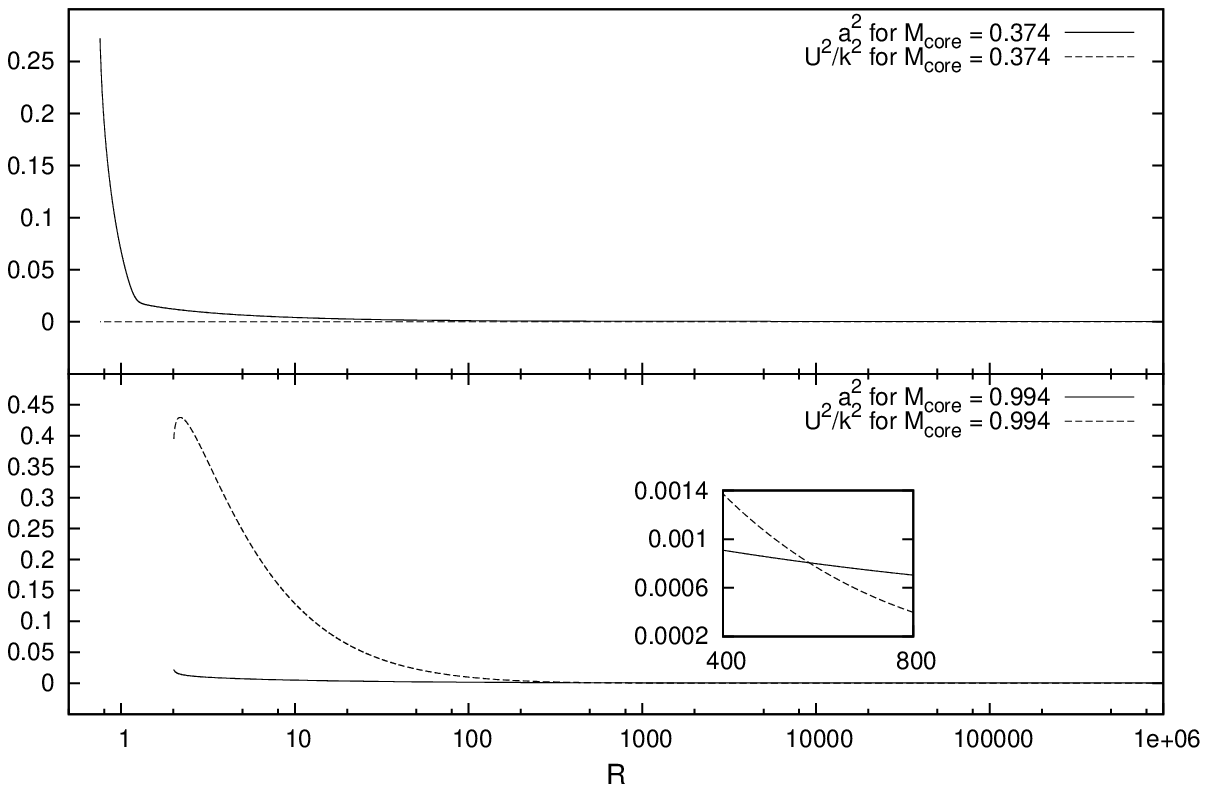}
\caption{\label{fig:4}High binding energy, $\alpha = 0.9$. A pair of extreme
subsonic and transonic flows corresponding to the same boundary data $L_0 = 0.1 L_E$ and
$a^2_\infty = 0.0004$ and different baryonic densities.
 Abscissa shows squares of the speed of sound and infall
velocity and the ordinate shows the radius $R$. A smaller picture inlet 
in the lower figure 
shows the vicinity of the sonic point.}
\end{figure}

Two forthcoming figures demonstrate that with increase of the parameter
$\alpha$
the shape of the set of subsonic solutions does not change significantly.
Its
boundary, however, can consist both of subsonic or transonic flows. We
call
corresponding solutions lying on the bifurcation curve as extreme.
Figure \ref{fig:2} is done for $\alpha = 0.5$.
Figure \ref{fig:3} presents the bifurcation curve for $\alpha = 0.9$. The
bifurcation point with the luminosity $L_0=0.31130L_E$ on the extreme curve
corresponds to a subsonic solution (see Fig.\ \ref{fig:8} for the
behaviour
of its speed of sound and of the infall velocity).
A transonic solution exists for $L_0=0.1 L_E$, $\alpha = 0.9$ on the
lighter
branch, but the partner lying on the more massive branch is a subsonic
flow, as
shown in Fig.\ \ref{fig:4}.

\begin{figure}[h]
\includegraphics[width=\linewidth ]{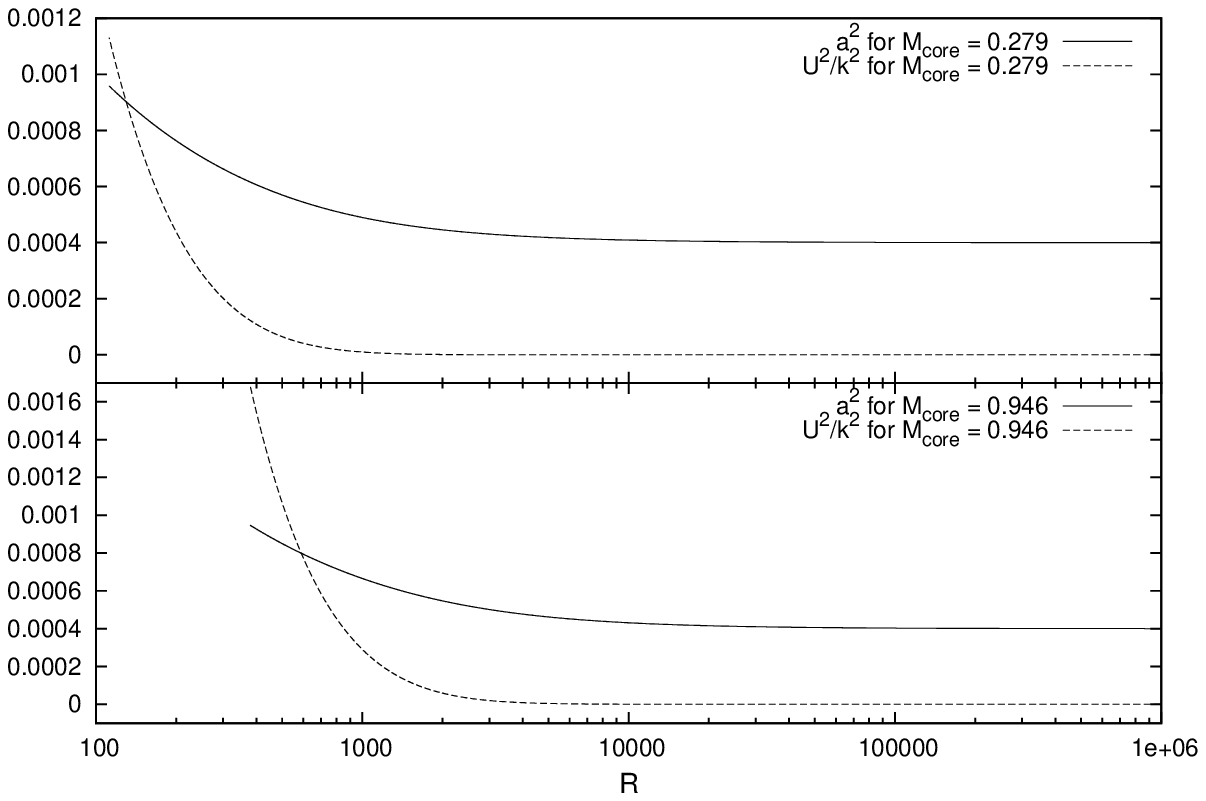}
\caption{\label{fig:5} $\alpha =2.5\times 10^{-3}$, $L_0/L_E=0.00318008$.
Behaviour of $a^2$ and $U^2/k^2$ in two supersonic
flows generated by the same boundary data but different baryonic denisties.}
\end{figure}

Figure \ref{fig:5} shows how squares of the speed of sound $a^2$ and of
the
spatial velocity $U^2/k^2$ depend on $R$ for different gas abundances. The
flow
with greater gas abundance (hence with a relatively lighter compact
center)
possesses a sonic point that is closer to the center than in the other
case.
Intuitive explanation is that for greater gas density the radiation
pressure is
bigger and prohibits quick falloff; the infall velocity can approach the
speed
of sound only close to the gravity center.

\begin{figure}[h]
\includegraphics[width=\linewidth ]{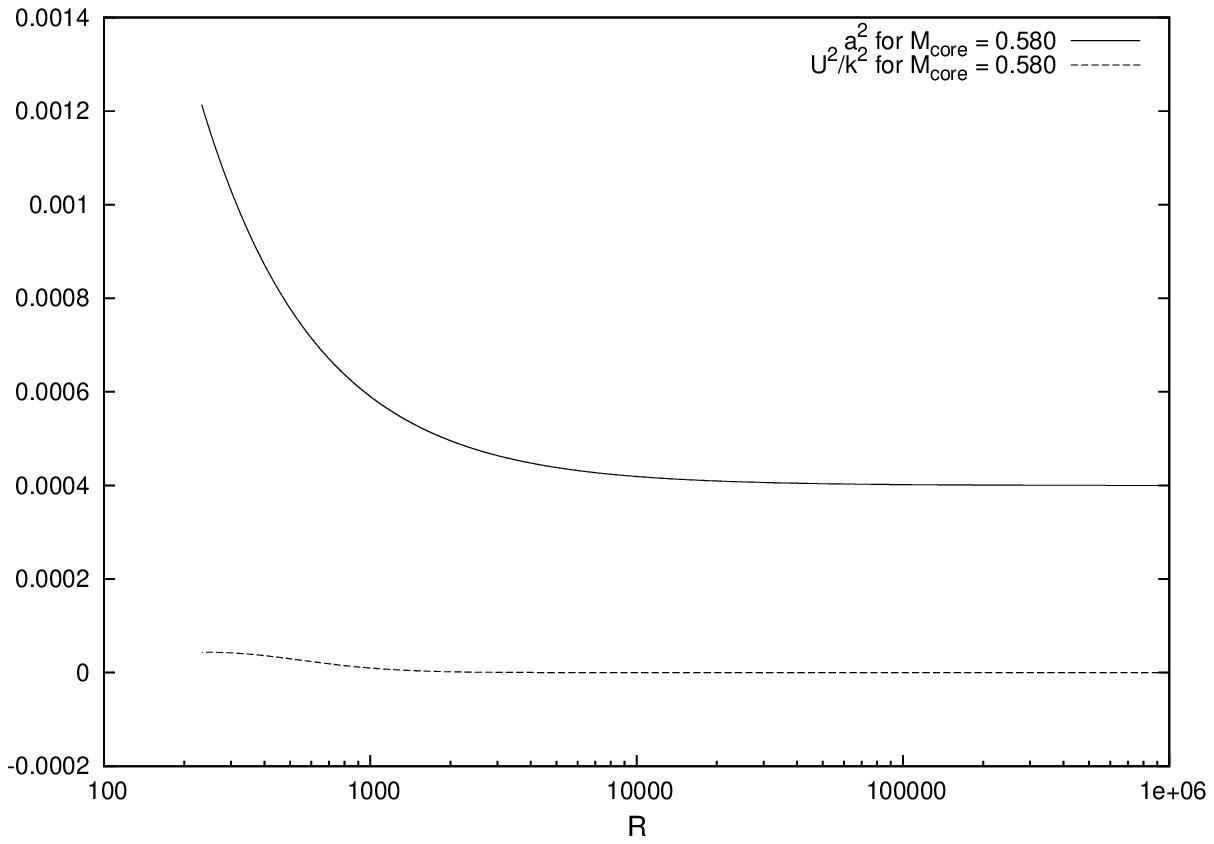}
\caption{\label{fig:6}  $\alpha =2.5\times 10^{-3}$, $L_0/L_E=0.00318008$.
 Behaviour of $a^2$ and $U^2/k^2$ in a subsonic flow
generated by the same boundary data as in Fig. \ref{fig:5}.}
\end{figure}

Figure \ref{fig:6} sketches the behaviour of the speed of sound and
$U^2/k^2$
for one of the subsonic solutions, with identical boundary  data -- but different baryonic densities
--  as for the
flows depicted in Fig.\ \ref{fig:5}.

The next figure shows characteristics of the bifurcation solution for the
binding energy parameter $\alpha = 0.0025$. The striking feature is that
the content of  gas abundance approaches $0.31$. We observed earlier that in
the
pure hydrodynamic accretion \cite{PRD2006} the maximum accretion rate
occurs
when the gas abundance is equal to $1/3$. That is valid both in the
general-relativistic \cite{PRD2006} and newtonian \cite{Kinasiewicz2007}
case.
In the Shakura model \cite{AA} one can analytically prove that at the
bifurcation point the luminosity is maximal and the gas abundance must be
smaller that $1/3$. The inspection of Figures 1 -- 3 shows that  this property
is satisfied also in the general-relativistic case. 
Here brightest flows coincide with bifurcation points. Their  gas abundance
is always  smaller than 1/3 and decreases with the increase of $L_0/L_E$.
\begin{figure}[h]
\includegraphics[width=\linewidth]{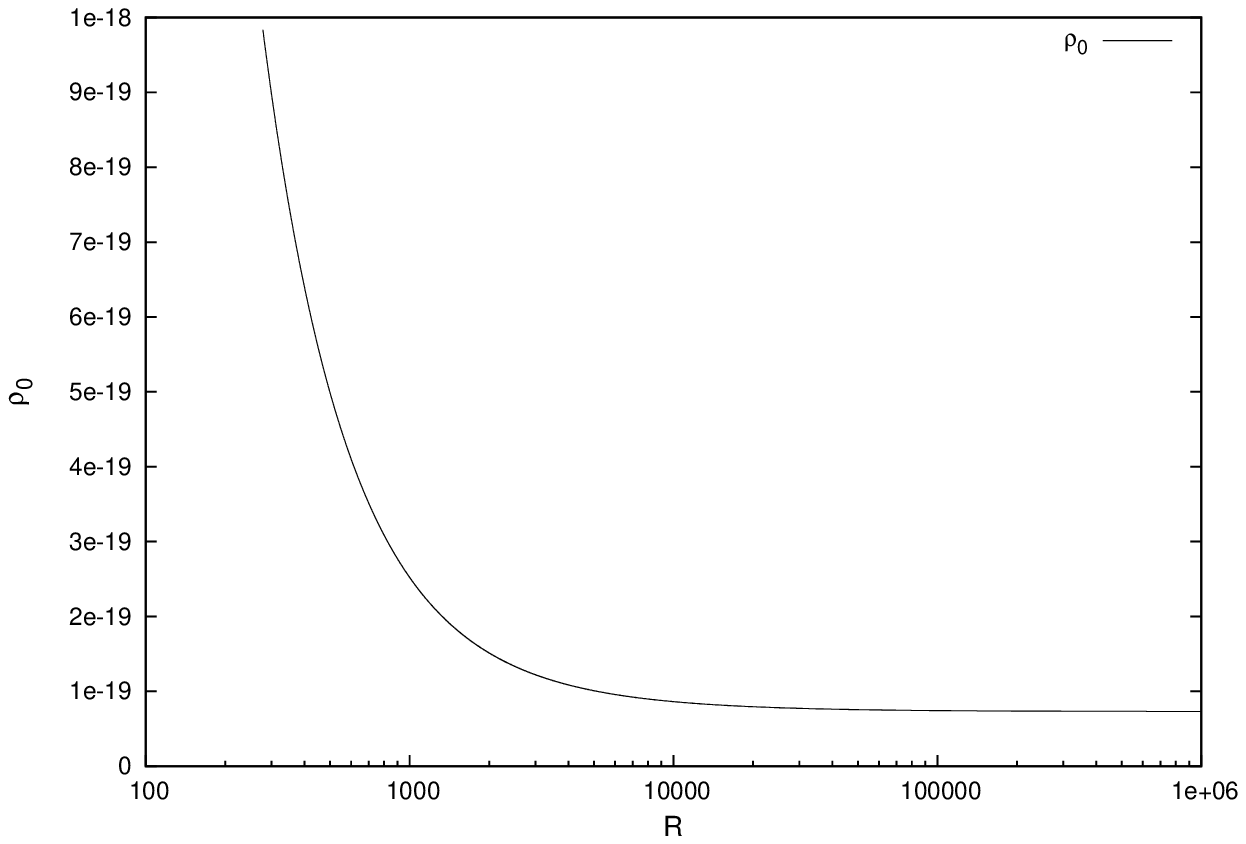}
\caption{\label{fig:7}  $\alpha =2.5\times 10^{-3}$, $L_0/L_E=0.01036, M_{core}=0.6933$.
    The baryonic mass density of the bifurcation solution
  in function of $R$.}
\end{figure}
The last figure displays the dependence of $a^2$ and $U^2/k^2$ on the area
radius $R$ in the case of the most luminous solutions corresponding to
$\alpha = 0.9$. It is clear that it is a subsonic flow.
\begin{figure}[h]
\includegraphics[width= \linewidth ]{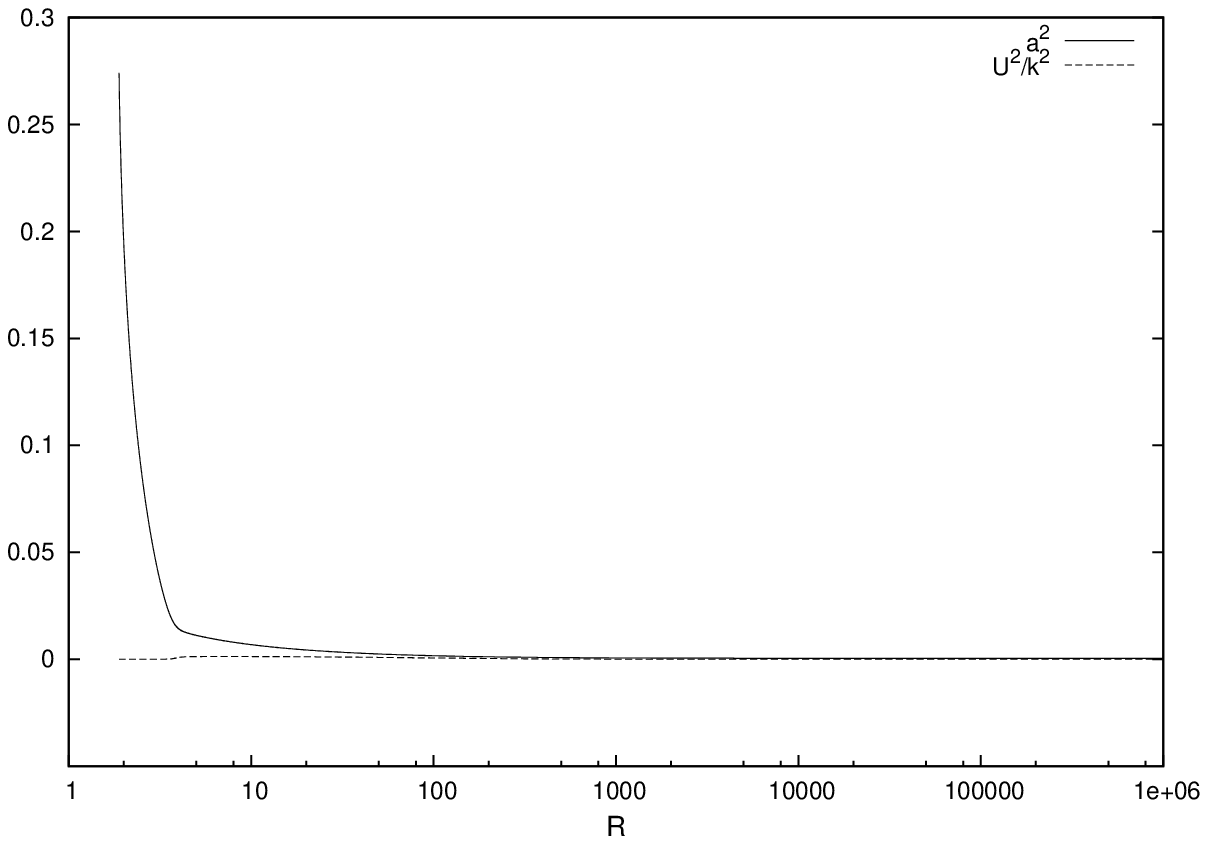}
\caption{\label{fig:8}  $\alpha =0.9$, $L_0/L_E=0.31130, M_{core}=0.932$.
 $a^2$ and $U^2/k^2$ in function of $R$.}
\end{figure}

\section{Concluding remarks.}

The present analysis is fully general-relativistic, with the backreaction
effects included. The   picture emerging here  is different from the former investigation
of spherical accretion in which the space-time geometry has been fixed
and therefore backreaction has been ignored \cite{inni}. The main new 
feature is the existence of an upper limit for the asymptotic baryonic mass
density and of a massive bifurcation branch. The new striking 
element is the fact that the brightest configuration is unique
for given redshift $\alpha $ and asymptotic speed of sound. In the case of 
low $\alpha $ (which is associated with low luminosity) the brightest object
flow is rich in gas --- about 1/3 of its mass is in the gaseous accreting matter
\cite{PRD2006}.

On the other hand  general-relativistic  radiating systems with
accreting gas behave in a qualitatively similar way to selfgravitating 
newtonian ones for small
redshifts (\textsl{i.e.}, for small binding energies). In both cases one
observes the following feature: two arms of the bifurcation curve of
transonic flows
embrace, in the diagram $L_0-m(R_0)$ (or the gas abundance versus
asymptotic
luminosity $L_0$), a set of subsonic solutions (see Fig.\ \ref{fig:1}).
For
given asymptotic data ($M$ --- the total mass, $L_0$ --- asymptotic luminosity and $a^2_\infty $ ---
the
asymptotic speed of sound) there exist two transonic solutions and an
infinite
number of subsonic solutions with intermediate baryonic densities. The
mass of
the central core in subsonic flows is comprised between two bounds of the
two
limiting transonicflows. The bifurcation point, where the two supersonic
branches cross and the luminosity is maximal for a given (binding energy)
parameter $\alpha$, is unique.

Some features of this picture change in the case of larger redshifts. This
happens at some value of the parameter $\alpha$ larger than $10^{-2}$. The
massive part of the bifurcation curve is  replaced by a curve of
subsonic solution (inspect Figs \ref{fig:2} and \ref{fig:3}). Transonic
flows survive only  on that section of the test fluid bifurcation branch
where  most of the mass is comprised in the compact core and  the luminosity is relatively
low. One observes (for instance for $\alpha = 0.5$ and $\alpha =0.9$) that as one
increases the total luminosity $L_0$, the transonic flows completely cease to exist on the
bifurcation curve and they are replaced by subsonic solutions. Nevertheless the set of all
flows has a similar shape that in the case of small $\alpha$, as exemplified in
Fig.\ \ref{fig:1} --  \ref{fig:3}. Subsonic solutions are not
specified uniquely for given boundary data, as we point out earlier, but
the
length of the interval of allowed values of the asymptotic baryonic
density
$\rho_{0\infty }$ becomes shorter with the increase of $L_0$. The solution
corresponding to the maximal luminosity is unique. In particular flows
corresponding to highest possible luminosities (that can be close to the
Eddington luminosity $L_E$ if $\alpha$ is close to $1$) are uniquely
determined.

Finally, it is interesting that quasi-stationary solutions of the model
considered in this paper can have a significant abundance of the gas.
Accreting
systems with maximal luminosities (in particular close to the Eddington
luminosity) can possess even 33\% of gas for small redshifts and still
almost  10\% of gas for $\alpha =0.9$. It is an open and important question
whether this picture is valid for generic nonspherical flows.

\section*{Acknowledgments.}
This paper has been partially supported by the MNII grant 1PO3B 01229.
Zdobys\l aw \'Swierczy\'nski thanks the Pedagogical University for the
research
grant. Krzysztof Roszkowski thanks the Foundation for Polish Science for
financial support.

\end{document}